\documentclass[sigconf,nonacm=true]{acmart}

\usepackage{pdfpages}
\usepackage{amsfonts}
\usepackage{pifont}
\usepackage{colortbl}
\usepackage{xcolor}
\usepackage{tcolorbox}
\usepackage{microtype}
\usepackage{multirow, booktabs}
\usepackage{adjustbox}
\usepackage{enumitem}
\usepackage{float}
\usepackage{bm}
\usepackage{bbm} 
\usepackage{amsmath}
\usepackage{graphicx}
\usepackage{hyperref}
\usepackage{soul}
\usepackage[labelformat=simple]{subcaption}
\usepackage{color}
\usepackage{fancyhdr}

\usepackage[framemethod=TikZ]{mdframed}
\usepackage{listings}
\newmdenv[
    innertopmargin=3pt,%
    linecolor=blue!20,%
    linewidth=1pt,%
    topline=true,%
    frametitleaboveskip=\dimexpr-\ht\strutbox\relax%
]{TasksBox}

\NewDocumentEnvironment{Tasks}{ m +b }{%
    \begin{TasksBox}[frametitle=\TaskTitle{#1}]\relax%
        #2
    \end{TasksBox}%
}{}

\NewDocumentCommand{\TaskTitle}{ m }{%
    \tikz[
        baseline=(current bounding box.east),
        outer sep=0pt,
    ]
        \node[
            anchor=east,
            rectangle,fill=blue!10,
        ] {
            \strut%
            #1
        };%
}

\newcommand{\etal}{\textit{et al.}}

\begin{document}
\title{Exploring Fine-tuning ChatGPT for News Recommendation}

\author{Xinyi Li}
\affiliation{%
\institution{Northwestern University, IL, US}
\city{}
\state{}
\country{}
}
\email{xinyili2024@u.northwestern.edu}

\author{Yongfeng Zhang}
\affiliation{%
\institution{Rutgers University, NJ, US}
\city{}
\state{}
\country{}
}
\email{yongfeng.zhang@rutgers.edu}

\author{Edward C. Malthouse}
\affiliation{%
\institution{Northwestern University, IL, US}
\city{}
\state{}
\country{}
}
\email{ecm@northwestern.edu}
\begin{abstract}

News recommendation systems (RS) play a pivotal role in the current digital age, shaping how individuals access and engage with information. The fusion of natural language processing (NLP) and RS, spurred by the rise of large language models such as the GPT and T5 series, blurs the boundaries between these domains, making a tendency to treat RS as a language task. ChatGPT, renowned for its user-friendly interface and increasing popularity, has become a prominent choice for a wide range of NLP tasks. While previous studies have explored ChatGPT on recommendation tasks, this study breaks new ground by investigating its fine-tuning capability, particularly within the news domain. In this study, we design two distinct prompts: one designed to treat news RS as the ranking task and another tailored for the rating task. We evaluate ChatGPT's performance in news recommendation by eliciting direct responses through the formulation of these two tasks. More importantly, we unravel the pivotal role of fine-tuning data quality in enhancing ChatGPT's personalized recommendation capabilities, and illustrates its potential in addressing the longstanding challenge of the ``cold item'' problem in RS. Our experiments, conducted using the Microsoft News dataset (MIND), reveal significant improvements achieved by ChatGPT after fine-tuning, especially in scenarios where a user's topic interests remain consistent, treating news RS as a ranking task. This study illuminates the transformative potential of fine-tuning ChatGPT as a means to advance news RS, offering more effective news consumption experiences.
\end{abstract}

\ccsdesc[500]{Information systems~Recommender systems}
\keywords{ChatGPT, Large Language Models, News Recommendations, Recommendation Systems, Information Retrieval.}

\maketitle

\section{Introduction}

In today's information-rich society, the accessibility of online news platforms Google News and Microsoft News has surged, offering users a vast array of news articles for consumption \cite{wu2020mind}. However, the sheer daily volume of new news articles presents a challenge for users seeking content aligned with their interests \cite{lian2018towards}. To address this issue, news RS play a crucial role in helping users discover articles relevant to their preferences. By effectively tailoring news recommendations, these systems not only enhance the user experience but also play a pivotal role in ensuring that individuals remain well-informed and engaged in a world inundated with information.

In the realm of news RS, models designed to comprehend article content and user interests are vital for delivering relevant recommendations. Techniques like the Gated Recurrent Unit (GRU) \cite{cho2014learning}, Long-Short Term Memory (LSTM) \cite{staudemeyer2019understanding}, Convolutional Neural Networks (CNNs) \cite{chen2015convolutional}, and attention mechanisms \cite{vaswani2017attention} have been popular choices for modeling user interests and comprehending article content \cite{an2019neural, wu2022news, wu2019neural}. However, these existing models are trained from scratch and may necessitate architectural modification when additional information is introduced. In response to these challenges, recent studies have shifted their focus toward using pre-trained language models. To leverage the pre-trained language models, researchers have introduced the concept of prompt learning \cite{jin2021good}, where specific prompts guide the output generation. Prompt learning makes it possible to generate outputs that adapt to the input and has been an effective approach for various NLP tasks \cite{jin2021good}, prompting researchers in the RS domain to recognize the potential of treating recommendation as a language task, harnessing the power of these techniques \cite{cui2022m6, geng2022recommendation,xu2023openp5}. 

ChatGPT, developed by OpenAI, has recently attracted substantial attention for its remarkable performance in various NLP tasks. While some preliminary studies have been conducted to explore its potential in recommendation tasks \cite{zhang2023chatgpt, li2023preliminary, liu2023chatgpt, bang2023multitask}, OpenAI's decision to allow fine-tuning of ChatGPT through their provided API represents an uncharted territory in research. This fine-tuning capability, offering the potential to enhance ChatGPT's performance, has yet to be examined.

To bridge the research gap, this study explores using ChatGPT to improve personalized news recommendations through fine-tuning, capitalizing on its linguistic capabilities. Specifically, our study entails the fine-tuning of ChatGPT by formulating the news recommendation as direct ranking and rating tasks. Furthermore, we delve into the critical role played by the quality of fine-tuning data in augmenting ChatGPT's capability in delivering better recommendations. Our experiments, conducted on the MIND dataset, reveal substantial improvements in ChatGPT's performance after fine-tuning, particularly when users maintain consistent topic interests. Additionally, our findings offer promising insights, indicating that fine-tuned performance surpasses certain established baselines when the proportion of ``cold'' items in the testing set falls below a certain threshold when treating news RS as a ranking task.

\section{Related Work}\label{section:literature}
\textbf{Sequential News Recommendation.} Sequential news recommendation methods are centered around predicting a user's preference for a candidate article based on their prior reading behavior. They play a critical role in delivering timely and relevant content to users in dynamic news environments. The wealth of textual information within news articles has prompted the application of language techniques to extract valuable insights and understand user interests \cite{an2019neural,wu2022news,wu2019neural}. For instance, Okura \etal\ \cite{okura2017embedding} introduced the use of a denoising autoencoder to analyze news representations and utilized a GRU network to model users' interests. An \etal\ \cite{an2019neural} adopted CNN and attention mechanisms to learn news representations from attributes such as title, topic, and subtopic. The NRMS model proposed by Wu \etal\ \cite{wu2019neural2} explored news representation from titles using a word-level, multi-head, self-attention mechanism and an additive word-attention network. In this work, instead of constructing models from scratch for news recommendation, we focus on leveraging pre-trained large language models (LLMs), specifically ChatGPT, to enhance news RS.

\textbf{Large Language Models and RS.} Pre-trained language models like BERT \cite{devlin2018bert} and GPT \cite{radford2018improving}, trained on extensive datasets, have demonstrated remarkable adaptability to various downstream tasks, and the integration of prompt learning techniques \cite{cho2014learning} has further enhanced their performance. This transformation has not been confined to NLP alone, it has also extended its reach to the realm of RS. Increasingly, recommendation tasks are being approached as language tasks. Researchers have proposed a multitude of innovative approaches in this context, including the conversion of item-based recommendation into text-based tasks \cite{geng2022recommendation}, the utilization of textual descriptions for understanding user behavior \cite{cui2022m6}, personalized prompt learning for explainable recommendation \cite{li2022personalized}, the learning of LLM-compatible IDs for precise generation, and the adoption of flexible multi-modality modeling methodologies for RS \cite{geng2023vip5}. LLMs and prompt learning techniques have also found their way into the field of news recommendation. For instance, Zhang \etal\ \cite{zhang2023prompt} employed prompt learning to address news recommendation by framing it as a slot filling task for [MASK] prediction, while Li \etal\ \cite{li2023pbnr} formulated news recommendation as a direct generative recommendation task using a pre-trained T5 \cite{raffel2020exploring} as the backbone.

ChatGPT has rapidly gained widespread popularity, prompting numerous studies to explore its capabilities and constraints. Qin \etal\ \cite{qin2023chatgpt} conducted an evaluation of ChatGPT's performance across a spectrum of NLP tasks, while Bang \etal\ \cite{bang2023multitask} comprehensively assessed its abilities in multitasking, multimodal applications, and multilingual contexts. On a parallel front, Liu \etal\ \cite{liu2023chatgpt} constructed a benchmark to evaluate ChatGPT's proficiency in various RS tasks, including rating prediction, sequential recommendation, direct recommendation, explanation generation, and review summarization. Dai \etal\ \cite{dai2023uncovering} conducted experiments to enhance ChatGPT's recommendation capabilities by aligning it with traditional information retrieval ranking capabilities, including point-wise, pair-wise, and list-wise methods. While previous studies have emphasized ChatGPT's zero-shot or few-shot capabilities for RS, in this paper, we aim to conduct a preliminary evaluation of ChatGPT's potential in news recommendation, uniquely positioned after fine-tuning, which involves customizing ChatGPT for news recommendation using the MIND dataset. Furthermore, we seek to uncover how the quality of fine-tuning data samples impact ChatGPT's efficacy for news recommendations.

\section{Recommendation Prompts}
A distinguishing feature of ChatGPT is its ability to yield impressive results when using the released model and subsequently fine-tuning it, particularly in cases where data is limited. In this section, we delve into the assessment of ChatGPT's recommendation capabilities, focusing on its performance after fine-tuning. To explore fine-tuned ChatGPT's suitability for news RS, we meticulously designed prompts tailored to two common and critical tasks in the RS domain: ranking and rating tasks.

\begin{figure*}
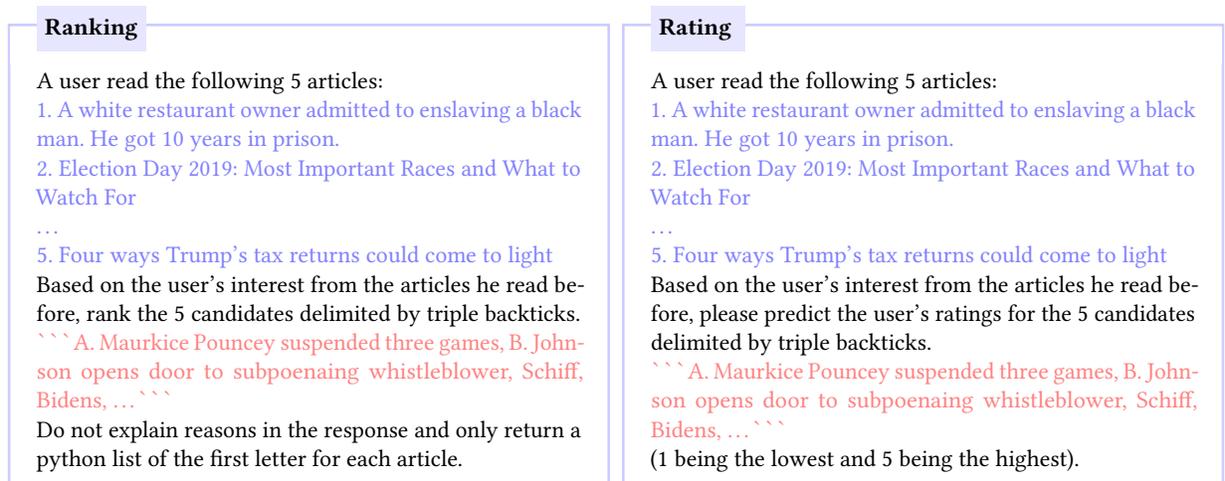

     \adjustbox{valign=t}{
        \begin{minipage}{0.45\textwidth}
            \begin{Tasks}{Ranking}
                A user read the following 5 articles: \\
                \textcolor{blue!50}{1. A white restaurant owner admitted to enslaving a black man. He got 10 years in prison.}\\
                \textcolor{blue!50}{2. Election Day 2019: Most Important Races and What to Watch For}\\
                \textcolor{blue!50}{\dots} \\
                \textcolor{blue!50}{5. Four ways Trump's tax returns could come to light}\\
                Based on the user's interest from the articles he read before, rank the 5 candidates delimited by triple backticks. \\
                \textcolor{red!50}{\textasciigrave \textasciigrave \textasciigrave A. Maurkice Pouncey suspended three games, B. Johnson opens door to subpoenaing whistleblower, Schiff, Bidens, \dots \textasciigrave \textasciigrave \textasciigrave } \\
                Do not explain reasons in the response and only return a python list of the first letter for each article.
            \end{Tasks}
        \end{minipage}%
    }
    \adjustbox{valign=t}{
        \begin{minipage}{0.45\textwidth}
            \begin{Tasks}{Rating }
                A user read the following 5 articles: \\
                \textcolor{blue!50}{1. A white restaurant owner admitted to enslaving a black man. He got 10 years in prison.}\\
                \textcolor{blue!50}{2. Election Day 2019: Most Important Races and What to Watch For}\\
                \textcolor{blue!50}{\dots} \\
                \textcolor{blue!50}{5. Four ways Trump's tax returns could come to light}\\
                Based on the user's interest from the articles he read before, please predict the user's ratings for the 5 candidates delimited by triple backticks.\\
                \textcolor{red!50}{\textasciigrave \textasciigrave \textasciigrave A. Maurkice Pouncey suspended three games, B. Johnson opens door to subpoenaing whistleblower, Schiff, Bidens, \dots \textasciigrave \textasciigrave \textasciigrave } \\
                (1 being the lowest and 5 being the highest).
            \end{Tasks}
        \end{minipage}
    }
\caption{Example prompts of both ranking and rating tasks, with the system content `You are a news recommender now'.}
\label{fig:prompt} 
    
\end{figure*}

\textbf{\textit{Ranking}.}
The ranking task in RS involves generating an ordered list of items for a user based on their preferences, historical interactions, or contextual information. The primary goal is to present the most relevant items at the top of the list to enhance the user's experience. In the context of our study, the ranking task is exemplified by the prompt shown in Figure~\ref{fig:prompt}. For a user denoted as $u \in \mathcal{U}$, we provide the articles that the user most recently interacted with $\{h_1, h_2, \dots \} \in \mathcal{I}$. Simultaneously, we also supply a list of candidate articles, denoted as $\{c_1, c_2, \dots\} \in \mathcal{I}$. The system is asked to directly sort these candidate articles based on the user's preference, which are analyzed from the user's past interactions with articles.


\textbf{\textit{Rating}.}
The rating task in RS is centered around the prediction of a rating score to a specific item for a user and this task is prevalent in scenarios where users explicitly rate items, providing feedback on their preferences. In the standard rating task prompt we designed, shown in Figure~\ref{fig:prompt}, a user denoted as $u$ is presented with the articles he/she most recently read $\{h_1, h_2, \dots \} \in \mathcal{I}$, along with a list of candidate articles, denoted as $\{c_1, c_2, \dots\} \in \mathcal{I}$. 
We then instruct the system to directly predict the rating scores for the candidate articles. 
The rating scale employed ranges from 1 to 5, where 5 denotes the highest score and 1 represents the lowest score. The system is encouraged to provide rating scores by making comparisons among the candidate articles.


\section{Experiments}
In this section, we conduct experiments to assess the effectiveness of fine-tuning ChatGPT. Through the performance comparison, we aim to answer the following research questions:

\begin{itemize}
\item \textbf{RQ1:} How does the performance of fine-tuned ChatGPT compare to that of ChatGPT in a zero-shot setting and other baseline models?

\item \textbf{RQ2:} How does fine-tuned ChatGPT perform by prompting news RS as different tasks -- ranking and rating?

\item \textbf{RQ3:} How does the sample size used for fine-tuning affect the performance of fine-tuned ChatGPT?

\item \textbf{RQ4:} What properties of data samples used for fine-tuning affect the performance of fine-tuned ChatGPT?
\end{itemize}

\subsection{Experimental Settings}
\subsubsection{Dataset}
For our experimental studies, we utilize the MIND dataset \cite{wu2020mind}. News recommendation presents unique challenges compared to other domains, as it may not always be highly personalized, and the nature of news is characterized by rapid changes \cite{dai2023uncovering}. To comprehensively assess whether fine-tuning can enhance news recommendation performance, we conduct evaluations across two distinct groups of customers:

\begin{itemize}
\item \textbf{Group 1:} This group consists of 100 randomly selected customers whose clicked article in the impression aligns with the topics they have previously read, i.e., the clicked article is from a topic that they have previously read.
\item \textbf{Group 2:} This group consist of 100 randomly selected customers whose clicked article in the impression is from a different topic than those they have previously read.
\end{itemize}
The division of customers into these two groups allows us to capture the dual nature of news recommendation: personalized recommendation that align with user interests and the challenges of offering recommendations outside a user's established preferences. This division also enables evaluation to determine if fine-tuning ChatGPT could enhance news recommendation across diverse scenarios.

\subsubsection{Baselines}
We compare the performance of fine-tuned ChatGPT with the following baseline models:
\begin{itemize}
\item \textbf{NAML} \cite{wu2022news}: models users' and articles' representations via multi-view self-attention. 
\item \textbf{LSTUR} \cite{an2019neural}: captures a user's interests by modeling both his long- and short-term preferences.
\item \textbf{NRMS} \cite{wu2019neural2}: models users' and articles' representations via multi-head self-attention network.
\item \textbf{Popularity} \cite{ji2020re}: recommends the
top-$k$ popular articles. 
\item \textbf{Zero-shot}: recommends the top-$k$ articles from the candidate pool, using ChatGPT's zero-shot capabilities.

\end{itemize}

\subsubsection{Metrics}
In numerical evaluations, we adopt metrics top-$k$ Normalized Discounted Cumulative Gain (NDCG@$k$) and Mean Reciprocal Rank (MRR@$k$) to assess the news recommendation performance.

\subsubsection{Implementation Details.} 
We evaluate using ChatGPT for news recommendation using \textit{gpt-3.5-turbo} for fine-tuning and zero-shot experiments. 
It is noteworthy that when utilizing zero-shot performance, the output generated by ChatGPT may not always adhere to the desired format requirements. To ensure compliance with format requirements and to meet the criteria for evaluation, a regeneration approach is employed, iteratively generating responses until the required format is met. Furthermore, in the context of the rating task within the zero-shot setting, where diverse rating values are anticipated to reflect comparative preferences, an additional format requirement is introduced. This requirement instructs ChatGPT to predict distinct rating scores for various candidates.

Additionally, it's important to mention that the data samples used for fine-tuning remain consistent for ranking and rating tasks, separately for Group 1 and Group 2 customers. This consistency is crucial in ensuring fair and meaningful comparisons. The individuals in the training data do not overlap with those in the test data, although articles in the training set may appear in the test data. The fine-tuning epoch and other hyper-parameters are automatically selected by OpenAI based on the size of fine-tuning dataset. During the fine-tuning process, with a fixed prompt, fixed group, and a fixed fine-tuning sample size, we conduct five independent experiments using five independent training datasets. This approach evaluates the reliability of our findings.

\section{Performance Evaluations }
\begin{table*}
\resizebox{\textwidth}{!}{
\begin{tabular}{c|c|c|c|c|c|c|c|c}
\hline 
\multicolumn{1}{c|}{} &\multicolumn{4}{c|}{Group 1} & \multicolumn{4}{c}{Group 2}\\ 

\hline
Method & MRR@3 & MRR@5 & NDCG@3  & NDCG@5 & MRR@3 & MRR@5 & NDCG@3  & NDCG@5  \\
\hline
NAML & 0.4086 & 0.4882 & 0.4689 & 0.6136 & 0.3614 & 0.4599 & 0.4123 &  0.5917 \\
LSTUR & 0.4129 &  0.5041 & 0.4614 & 0.6256 & 0.4128 & 0.4978 & 0.4596 &  0.6213\\
NRMS & 0.4263 & 0.4984 & 0.4913 & 0.6219 & 0.3972 & 0.4911 & 0.4438 & 0.6151 \\

Popularity & 0.4764 & 0.5423 & 0.5355 & 0.6551 & \textbf{0.5264} & \textbf{0.5826} & \textbf{0.5829} & \textbf{0.6855} \\

\hline
Zero-shot (Ranking)  & 0.4446$\pm$0.0023 & 0.5258$\pm$0.0021 & 0.4936$\pm$0.0023  & 0.6420$\pm$0.0016 & 0.2935$\pm$0.0023 & 0.3930$\pm$0.0021 & 0.3604$\pm$0.0024 & 0.5415$\pm$0.0015 \\
Zero-shot (Rating) & 0.3735$\pm$0.0029 & 0.4540$\pm$0.0024 & 0.4428$\pm$0.0029 & 0.5886$\pm$0.0018

& 0.3754$\pm$0.0112 & 0.4691$\pm$0.0079 & 0.4266$\pm$0.0133 & 0.5984$\pm$0.0063\\



\hline
Fine-tuned (Ranking)& \textbf{0.5278$\pm$0.0719} & \textbf{0.5928$\pm$0.0598} & \textbf{0.5755$\pm$0.0682 } &\textbf{0.6930$\pm$0.0454 }  & 0.3802$\pm$0.0279 &  0.4690$\pm$0.0221 & 0.4372$\pm$0.0298 & 0.5989$\pm$0.0169  \\ 

Fine-tuned (Rating) & 0.3794$\pm$0.0249 & 0.4659$\pm$0.0223 & 0.4494$\pm$0.0234 & 0.5969$\pm$0.0168  & 0.3637$\pm$0.0209 & 0.4538$\pm$0.0199 & 0.4261$\pm$0.0245 & 0.5865$\pm$0.0145 \\

\hline
\end{tabular}}
\caption{The news recommendation performance on customers. Bold numbers indicate the best performance. 5 independent experiments are conducted for zero-shot ranking and rating, while 25 independent experiments are conducted for fine-tuning setting. The statistical significance was assessed using the Student’s t-test, with a significance level of $p$ < 0.05. }
\label{tab:result}
\end{table*}

\subsection{RQ1\&2: Performance Comparison}

Table~\ref{tab:result} presents the performance results for various models, including baseline models, zero-shot ChatGPT using the news RS ranking and rating task formulations, and fine-tuned ChatGPT with these same formulated tasks. We conduct separate evaluations for Group 1 and Group 2 customers, and here are our observations:

The first 4 baselines exhibit no variance, as they are intentionally trained with a significantly larger number of data samples than the fine-tuning sample sizes, aiming at establishing them as performance upper bounds for a more rigorous and superior baseline comparison. The popularity baseline stands out as a strong baseline for news recommendation, which is in line with the findings of many other research works \cite{dai2023uncovering, qi2021pp}. It consistently outperforms the zero-shot ChatGPT and other deep neural-based models for both Group 1 and Group 2 users. This is particularly evident when readers have engaged with articles from diverse topics. These findings underscore the distinct nature of news recommendation, where user behavior may not always align closely with personalized recommendations, as seen in other domains like e-commerce \cite{jonnalagedda2016incorporating, yang2016effects}. 

\begin{figure}[h]
  \centering  
 \includegraphics[width=0.9\columnwidth]{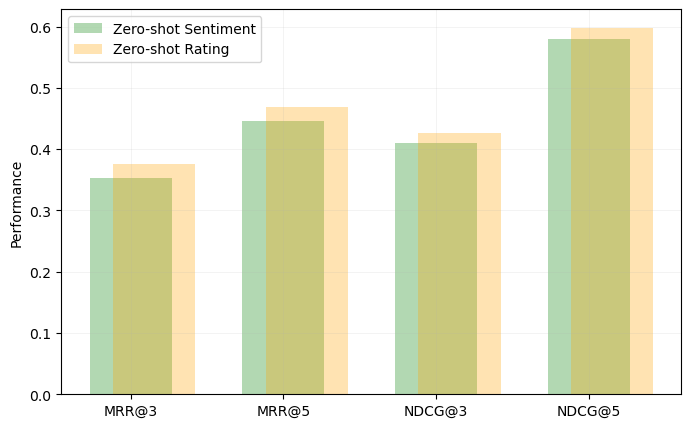}
  \caption{Comparison of zero-shot ChatGPT performance between sentiment classification for candidate article and the rating task among Group 2 customers. Five independent experiments are conducted, and the figure shows the average performances.}
  \label{fig:sentiment}
\end{figure}

In the zero-shot setup, ChatGPT's performances lag behind that of popularity-based models. When users' topic interests change, as observed in Group 2, ChatGPT's zero-shot performance using ranking task formulation falls short of all baseline models. This suggests that ChatGPT's strength lies in semantic understanding and its tendency to recommend articles similar to those previously read by users. However, an intriguing finding is that for the rating task the zero-shot performance on Group 1 and Group 2 customers are similar to each other. One possible explanation is that, within the zero-shot setup, ChatGPT interprets the rating task in a manner akin to sentiment classification, where a rating of 1 represents strongly negative and 5 indicates strongly positive. To substantiate this hypothesis, we introduce a similar prompt as the rating task, instructing ChatGPT to generate relative sentiment scores for candidate articles directly. The resulting zero-shot performance, treated as a sentiment classification task, is illustrated in Figure~\ref{fig:sentiment}. Our findings provide empirical support for our hypothesis that within the zero-shot context, ChatGPT perceives the rating task as a form of sentiment classification for candidates, a perspective that exhibits notable zero-shot performance as demonstrated in previous work \cite{wang2023chatgpt}. This interpretation
results in the noteworthy performance observed with Group 2 customers, and the observation also demonstrates the importance of proper prompt-based task formulation when fine-tuning ChatGPT for downstream tasks.

Under the fine-tuning setup, there is a notable improvement in performance compared to zero-shot with ranking task, particularly in Group 1. This improvement may be attributed to the fact that fine-tuning not only enhances ChatGPT's semantic understanding but also makes more effective use of position information. During the fine-tuning process, the clicked articles are consistently placed at the first position in the generated ranking list response, regardless of their original position in the provided candidate list. This allows the model to better exploit the positional information. In contrast, for the rating prompt, the five scores may manifest at various positions within the generated responses. For customers in Group 2, fine-tuning also leads to improvements compared to zero-shot, albeit not as substantial as observed in Group 1, even when using the same fine-tuning sample sizes. One possible explanation is that ChatGPT tends to recommend articles from diverse topics after fine-tuning. However, within the provided candidate articles, multiple options may satisfy this diversity requirement. Without knowledge of the popularity of these articles, ChatGPT might randomly select one to fulfill the diversity requirement.

However, under the fine-tuning setup, the performances are similar when using rating task, whether applied to Group 1 or Group 2, as compared to zero-shot approach. This might be attributed to the fact that, even during fine-tuning, while semantic understanding can be improved, the model's capacity for handling numerical comparison remains relatively unchanged. Additionally, the rating task lacks the advantage of utilizing positional information from the generated responses during fine-tuning, unlike the ranking task. Furthermore, the rating prompt necessitates the assignment of scores for all candidate simultaneously, making the rating task more challenging. Lastly, it's worth noting that different customers may use the rating scale differently (i.e., the system must learn user biases). This finding that ranking outperforms the rating task aligns with prior research, particularly comparing point-wise and list-wise ranking \cite{dai2023uncovering}. 


\subsection{RQ3: Performance Under Different Fine-tuning Sample Sizes}
\begin{figure}[h]
  \centering
  \begin{subfigure}[b]{0.45\textwidth}
    \centering
    \includegraphics[width=\textwidth]{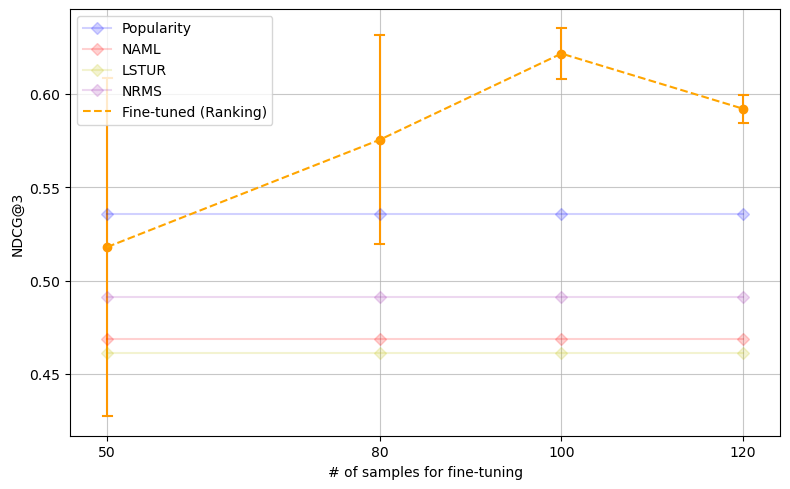}
  \end{subfigure}
  \hfill
  \begin{subfigure}[b]{0.45\textwidth}
    \centering
    \includegraphics[width=\textwidth]{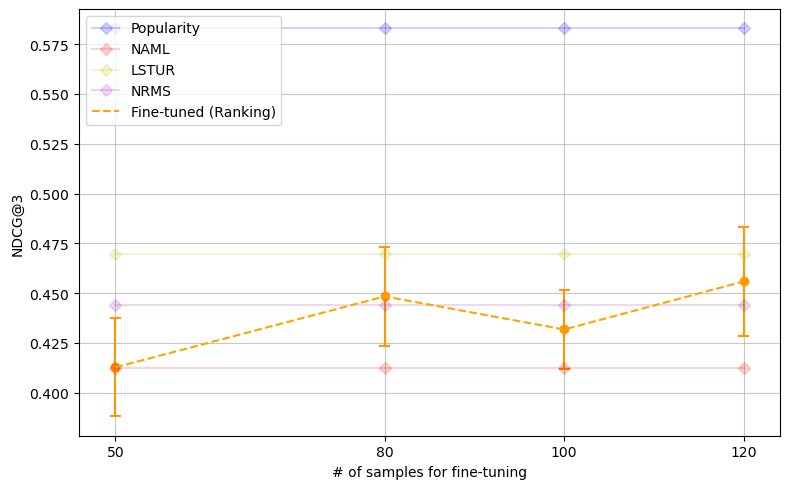}
  \end{subfigure}
  \caption{Recommendation performances with different quantities of fine-tuning samples. The first subfigure is for Group 1 while the second is for Group 2 readers.}
  \label{fig:num_finetune}
\end{figure}

Our experiments reveal a notable performance enhancement in ChatGPT when using ranking tasks after fine-tuning. In this subsection, we investigate how the quantity of fine-tuning samples affects fine-tuned ChatGPT's recommendation performance using ranking task. We conduct experiments varying the sample size within the range \{50, 80, 100, 120\}. For each sample size, fine-tuning is performed independently five times, utilizing distinct fine-tuning data samples. The results are presented in Figure~\ref{fig:num_finetune}, demonstrating the performance differences across sample sizes, as measured by NDCG@3.

Our observations indicate that the average performance (i.e., NDCG@$k$) remains consistent across different fine-tuning sample sizes, suggesting that the quantity of fine-tuned data samples does not significantly affect the performance of fine-tuned ChatGPT (the $p$-value from a one-way ANOVA, testing the equality of means, exceeds 0.1). Additionally, when the same number of samples was used for fine-tuning, there was variability in performance on the same test set. This not only reaffirms that performance is not solely contingent on the fine-tuning sample sizes but also emphasizes our interest in identifying the quality of data samples that enhance performance.

\subsection{RQ4: Quality of Fine-tuning Samples}
ChatGPT, when using ranking tasks after fine-tuning, even outperforms the popularity-based model for Group 1 users. In this subsection, we delve deeper into the realm of ranking tasks and aim to detect specific factors that boost fine-tuned ChatGPT's performance in news recommendation.

\begin{figure}[]
  \centering
  \begin{subfigure}[b]{0.45\textwidth}
    \centering
    \includegraphics[width=\textwidth]{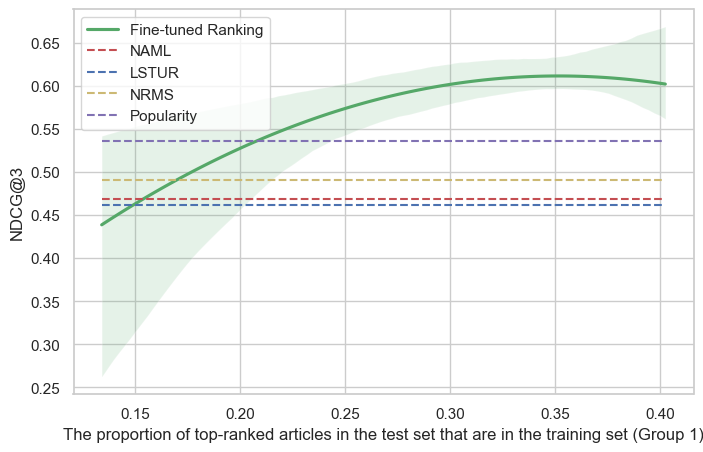}
  \end{subfigure}
  \begin{subfigure}[b]{0.45\textwidth}
    \centering
    \includegraphics[width=\textwidth]{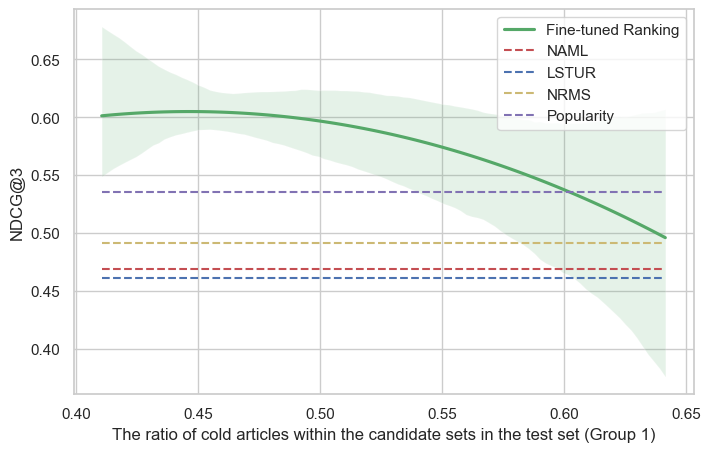}
  \end{subfigure}
  \begin{subfigure}[b]{0.45\textwidth}
    \centering
    \includegraphics[width=\textwidth]{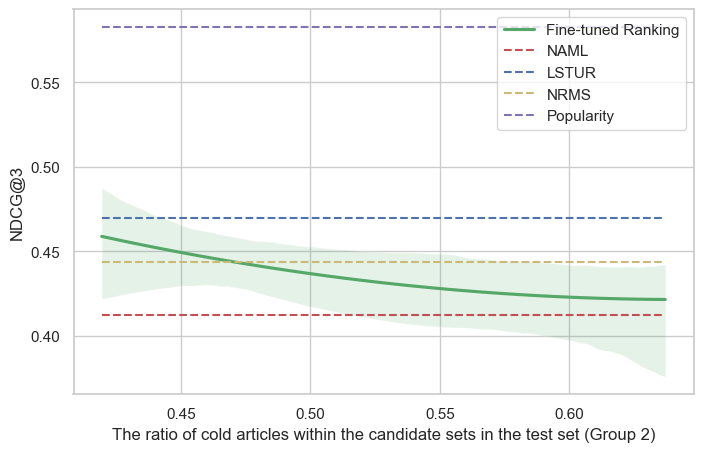}
  \end{subfigure}
  \caption{Influence of overlap ratio and ``cold'' articles on fine-tuned ChatGPT's news recommendation performance using the ranking prompt. Fine-tuned ChatGPT outperforms all baselines for Group 1 readers when the ratio of ``cold'' articles < 0.6, and surpasses the NAML baseline for Group 2 readers when the ratio < 0.45. }
  \label{fig:factor}
\end{figure}

The intriguing observation that fine-tuned ChatGPT, using ranking tasks, can even outperform the popularity-based model for Group 1 users motivates us to analyze the impact of the proportion of top-ranked articles in the test set that were also present in the training set. A higher proportion indicates more overlap between the articles users engaged with during training and testing. The first subfigure in Figure~\ref{fig:factor} illustrates that fine-tuned ChatGPT's performance for Group 1 customers shows improvement as the overlap ratio increases toward a certain threshold with statistical significance ($p$-value < 0.05). This finding may offer a possible explanation for the model's superior performance compared to the popularity-based model for Group 1 users. When ChatGPT encounters articles during testing that it has previously interacted with during the fine-tuning process, it might discern implicit popularity signals from these articles, utilizing the positional information derived from the ranking task. Group 1 users, with their consistent interests and ChatGPT's proficiency in textual understanding, benefit from this approach, leveraging both positional information and semantic understanding. Notably, this factor does not yield statistically significant effects for Group 2 users ($p$-value > 0.1).

We also investigate the impact of the presence of ``cold'' articles in the candidates during testing. A candidate article is labeled as ``cold'' if it is not part of the fine-tuning samples. As observed in the last two subfigures of Figure~\ref{fig:factor}, we find that the proportion of cold articles significantly influences fine-tuned ChatGPT's performance ($p$-values below 0.05 for Group 1 and below 0.1 for Group 2). In general, we notice that as the ratio of ``cold'' articles in the test set increases, the fine-tuned ChatGPT's performance decreases. The observation that fine-tuned ChatGPT can surpass specific baselines, which are trained with more data samples and fewer ``cold'' items during evaluation, underscores ChatGPT's potential in addressing the ``cold'' item challenge in RS, as long as the ratio of ``cold'' articles remains within a particular threshold, as shown in Figure~\ref{fig:factor}. 

\subsection{Computational Cost}
In our experiments, fine-tuning ChatGPT with a maximum of 120 samples typically took around 30 minutes to complete. This is done with an average of approximately 310 input tokens and using the default number of epochs once the fine-tuning process began. 

\section{Conclusion}
In this study, we conduct experiments that showcase the substantial benefits of fine-tuning ChatGPT for news recommendation. This may seem like a trivial task. However, as we have shown in the research, the performance of fine-tuning depends on several factors such as the topic alignment, prompt formulation, sample sizes, and the quality of fine-tuning samples. More specifically, we compare the effectiveness of ranking and rating tasks for fine-tuning ChatGPT, and our results indicate that ranking consistently outperforms rating by leveraging both positional cues from the generated responses during fine-tuning and semantic understanding. The challenges of rating tasks become evident as ChatGPT struggles with making numerical comparisons when tasked with generating ratings for all candidates simultaneously. Additionally, ChatGPT sometimes interprets the rating task as a sentiment classification task in the zero-shot mode, particularly for Group 2 customers. Moreover, we delve into the factors influencing ChatGPT's ranking performance after fine-tuning. Our investigation unveils the degree of overlap between the articles users interacted with during both training and testing is a significant factor when user interests remain consistent. One of the most promising findings in our study is ChatGPT's potential to address the ``cold'' item issue in RS. Despite competing with baselines trained on larger datasets with fewer ``cold'' items during evaluation, fine-tuned ChatGPT consistently outperforms them within a specific threshold of ratio of ``cold'' items. This observation underscores ChatGPT's capacity to mitigate the ``cold'' item issue to enhance RS. 

For future studies, we envision several promising research directions. Given the fundamental role of popularity in news recommendation, a notable avenue for future exploration is the effective incorporation of popularity-related information into prompts. Additionally, enhancing news recommendation for users when their interests undergo shifts, particularly via fine-tuning ChatGPT, holds significant potential for further advancement.

\bibliographystyle{ACM-Reference-Format}
\bibliography{ref}


\begin{thebibliography}{33}


\ifx \showCODEN    \undefined \def \showCODEN     #1{\unskip}     \fi
\ifx \showDOI      \undefined \def \showDOI       #1{#1}\fi
\ifx \showISBNx    \undefined \def \showISBNx     #1{\unskip}     \fi
\ifx \showISBNxiii \undefined \def \showISBNxiii  #1{\unskip}     \fi
\ifx \showISSN     \undefined \def \showISSN      #1{\unskip}     \fi
\ifx \showLCCN     \undefined \def \showLCCN      #1{\unskip}     \fi
\ifx \shownote     \undefined \def \shownote      #1{#1}          \fi
\ifx \showarticletitle \undefined \def \showarticletitle #1{#1}   \fi
\ifx \showURL      \undefined \def \showURL       {\relax}        \fi
\providecommand\bibfield[2]{#2}
\providecommand\bibinfo[2]{#2}
\providecommand\natexlab[1]{#1}
\providecommand\showeprint[2][]{arXiv:#2}

\bibitem[An et~al\mbox{.}(2019)]%
        {an2019neural}
\bibfield{author}{\bibinfo{person}{Mingxiao An}, \bibinfo{person}{Fangzhao Wu}, \bibinfo{person}{Chuhan Wu}, \bibinfo{person}{Kun Zhang}, \bibinfo{person}{Zheng Liu}, {and} \bibinfo{person}{Xing Xie}.} \bibinfo{year}{2019}\natexlab{}.
\newblock \showarticletitle{Neural news recommendation with long-and short-term user representations}. In \bibinfo{booktitle}{\emph{Proceedings of the 57th Annual Meeting of the Association for Computational Linguistics}}. \bibinfo{pages}{336--345}.
\newblock


\bibitem[Bang et~al\mbox{.}(2023)]%
        {bang2023multitask}
\bibfield{author}{\bibinfo{person}{Yejin Bang}, \bibinfo{person}{Samuel Cahyawijaya}, \bibinfo{person}{Nayeon Lee}, \bibinfo{person}{Wenliang Dai}, \bibinfo{person}{Dan Su}, \bibinfo{person}{Bryan Wilie}, \bibinfo{person}{Holy Lovenia}, \bibinfo{person}{Ziwei Ji}, \bibinfo{person}{Tiezheng Yu}, \bibinfo{person}{Willy Chung}, {et~al\mbox{.}}} \bibinfo{year}{2023}\natexlab{}.
\newblock \showarticletitle{A multitask, multilingual, multimodal evaluation of chatgpt on reasoning, hallucination, and interactivity}.
\newblock \bibinfo{journal}{\emph{arXiv preprint arXiv:2302.04023}} (\bibinfo{year}{2023}).
\newblock


\bibitem[Chen(2015)]%
        {chen2015convolutional}
\bibfield{author}{\bibinfo{person}{Yahui Chen}.} \bibinfo{year}{2015}\natexlab{}.
\newblock \emph{\bibinfo{title}{Convolutional neural network for sentence classification}}.
\newblock \bibinfo{thesistype}{Master's\ thesis}. \bibinfo{school}{University of Waterloo}.
\newblock


\bibitem[Cho et~al\mbox{.}(2014)]%
        {cho2014learning}
\bibfield{author}{\bibinfo{person}{Kyunghyun Cho}, \bibinfo{person}{Bart Van~Merri{\"e}nboer}, \bibinfo{person}{Caglar Gulcehre}, \bibinfo{person}{Dzmitry Bahdanau}, \bibinfo{person}{Fethi Bougares}, \bibinfo{person}{Holger Schwenk}, {and} \bibinfo{person}{Yoshua Bengio}.} \bibinfo{year}{2014}\natexlab{}.
\newblock \showarticletitle{Learning phrase representations using RNN encoder-decoder for statistical machine translation}.
\newblock \bibinfo{journal}{\emph{arXiv preprint arXiv:1406.1078}} (\bibinfo{year}{2014}).
\newblock


\bibitem[Cui et~al\mbox{.}(2022)]%
        {cui2022m6}
\bibfield{author}{\bibinfo{person}{Zeyu Cui}, \bibinfo{person}{Jianxin Ma}, \bibinfo{person}{Chang Zhou}, \bibinfo{person}{Jingren Zhou}, {and} \bibinfo{person}{Hongxia Yang}.} \bibinfo{year}{2022}\natexlab{}.
\newblock \showarticletitle{M6-Rec: Generative Pretrained Language Models are Open-Ended Recommender Systems}.
\newblock \bibinfo{journal}{\emph{arXiv preprint arXiv:2205.08084}} (\bibinfo{year}{2022}).
\newblock


\bibitem[Dai et~al\mbox{.}(2023)]%
        {dai2023uncovering}
\bibfield{author}{\bibinfo{person}{Sunhao Dai}, \bibinfo{person}{Ninglu Shao}, \bibinfo{person}{Haiyuan Zhao}, \bibinfo{person}{Weijie Yu}, \bibinfo{person}{Zihua Si}, \bibinfo{person}{Chen Xu}, \bibinfo{person}{Zhongxiang Sun}, \bibinfo{person}{Xiao Zhang}, {and} \bibinfo{person}{Jun Xu}.} \bibinfo{year}{2023}\natexlab{}.
\newblock \showarticletitle{Uncovering ChatGPT's Capabilities in Recommender Systems}.
\newblock \bibinfo{journal}{\emph{arXiv preprint arXiv:2305.02182}} (\bibinfo{year}{2023}).
\newblock


\bibitem[Devlin et~al\mbox{.}(2018)]%
        {devlin2018bert}
\bibfield{author}{\bibinfo{person}{Jacob Devlin}, \bibinfo{person}{Ming-Wei Chang}, \bibinfo{person}{Kenton Lee}, {and} \bibinfo{person}{Kristina Toutanova}.} \bibinfo{year}{2018}\natexlab{}.
\newblock \showarticletitle{Bert: Pre-training of deep bidirectional transformers for language understanding}.
\newblock \bibinfo{journal}{\emph{arXiv preprint arXiv:1810.04805}} (\bibinfo{year}{2018}).
\newblock


\bibitem[Geng et~al\mbox{.}(2022)]%
        {geng2022recommendation}
\bibfield{author}{\bibinfo{person}{Shijie Geng}, \bibinfo{person}{Shuchang Liu}, \bibinfo{person}{Zuohui Fu}, \bibinfo{person}{Yingqiang Ge}, {and} \bibinfo{person}{Yongfeng Zhang}.} \bibinfo{year}{2022}\natexlab{}.
\newblock \showarticletitle{Recommendation as Language Processing (RLP): A Unified Pretrain, Personalized Prompt \& Predict Paradigm (P5)}.
\newblock \bibinfo{journal}{\emph{RecSys}} (\bibinfo{year}{2022}).
\newblock


\bibitem[Geng et~al\mbox{.}(2023)]%
        {geng2023vip5}
\bibfield{author}{\bibinfo{person}{Shijie Geng}, \bibinfo{person}{Juntao Tan}, \bibinfo{person}{Shuchang Liu}, \bibinfo{person}{Zuohui Fu}, {and} \bibinfo{person}{Yongfeng Zhang}.} \bibinfo{year}{2023}\natexlab{}.
\newblock \showarticletitle{VIP5: Towards Multimodal Foundation Models for Recommendation}.
\newblock \bibinfo{journal}{\emph{EMNLP}} (\bibinfo{year}{2023}).
\newblock


\bibitem[Ji et~al\mbox{.}(2020)]%
        {ji2020re}
\bibfield{author}{\bibinfo{person}{Yitong Ji}, \bibinfo{person}{Aixin Sun}, \bibinfo{person}{Jie Zhang}, {and} \bibinfo{person}{Chenliang Li}.} \bibinfo{year}{2020}\natexlab{}.
\newblock \showarticletitle{A re-visit of the popularity baseline in recommender systems}. In \bibinfo{booktitle}{\emph{Proceedings of the 43rd International ACM SIGIR Conference on Research and Development in Information Retrieval}}. \bibinfo{pages}{1749--1752}.
\newblock


\bibitem[Jin et~al\mbox{.}(2021)]%
        {jin2021good}
\bibfield{author}{\bibinfo{person}{Woojeong Jin}, \bibinfo{person}{Yu Cheng}, \bibinfo{person}{Yelong Shen}, \bibinfo{person}{Weizhu Chen}, {and} \bibinfo{person}{Xiang Ren}.} \bibinfo{year}{2021}\natexlab{}.
\newblock \showarticletitle{A good prompt is worth millions of parameters? low-resource prompt-based learning for vision-language models}.
\newblock \bibinfo{journal}{\emph{arXiv preprint arXiv:2110.08484}} (\bibinfo{year}{2021}).
\newblock


\bibitem[Jonnalagedda et~al\mbox{.}(2016)]%
        {jonnalagedda2016incorporating}
\bibfield{author}{\bibinfo{person}{Nirmal Jonnalagedda}, \bibinfo{person}{Susan Gauch}, \bibinfo{person}{Kevin Labille}, {and} \bibinfo{person}{Sultan Alfarhood}.} \bibinfo{year}{2016}\natexlab{}.
\newblock \showarticletitle{Incorporating popularity in a personalized news recommender system}.
\newblock \bibinfo{journal}{\emph{PeerJ Computer Science}}  \bibinfo{volume}{2} (\bibinfo{year}{2016}), \bibinfo{pages}{e63}.
\newblock


\bibitem[Li et~al\mbox{.}(2022)]%
        {li2022personalized}
\bibfield{author}{\bibinfo{person}{Lei Li}, \bibinfo{person}{Yongfeng Zhang}, {and} \bibinfo{person}{Li Chen}.} \bibinfo{year}{2022}\natexlab{}.
\newblock \showarticletitle{Personalized prompt learning for explainable recommendation}.
\newblock \bibinfo{journal}{\emph{arXiv preprint arXiv:2202.07371}} (\bibinfo{year}{2022}).
\newblock


\bibitem[Li et~al\mbox{.}(2023a)]%
        {li2023pbnr}
\bibfield{author}{\bibinfo{person}{Xinyi Li}, \bibinfo{person}{Yongfeng Zhang}, {and} \bibinfo{person}{Edward~C Malthouse}.} \bibinfo{year}{2023}\natexlab{a}.
\newblock \showarticletitle{PBNR: Prompt-based News Recommender System}.
\newblock \bibinfo{journal}{\emph{arXiv preprint arXiv:2304.07862}} (\bibinfo{year}{2023}).
\newblock


\bibitem[Li et~al\mbox{.}(2023b)]%
        {li2023preliminary}
\bibfield{author}{\bibinfo{person}{Xinyi Li}, \bibinfo{person}{Yongfeng Zhang}, {and} \bibinfo{person}{Edward~C Malthouse}.} \bibinfo{year}{2023}\natexlab{b}.
\newblock \showarticletitle{A Preliminary Study of ChatGPT on News Recommendation: Personalization, Provider Fairness, Fake News}.
\newblock \bibinfo{journal}{\emph{arXiv preprint arXiv:2306.10702}} (\bibinfo{year}{2023}).
\newblock


\bibitem[Lian et~al\mbox{.}(2018)]%
        {lian2018towards}
\bibfield{author}{\bibinfo{person}{Jianxun Lian}, \bibinfo{person}{Fuzheng Zhang}, \bibinfo{person}{Xing Xie}, {and} \bibinfo{person}{Guangzhong Sun}.} \bibinfo{year}{2018}\natexlab{}.
\newblock \showarticletitle{Towards Better Representation Learning for Personalized News Recommendation: a Multi-Channel Deep Fusion Approach.}. In \bibinfo{booktitle}{\emph{IJCAI}}. \bibinfo{pages}{3805--3811}.
\newblock


\bibitem[Liu et~al\mbox{.}(2023)]%
        {liu2023chatgpt}
\bibfield{author}{\bibinfo{person}{Junling Liu}, \bibinfo{person}{Chao Liu}, \bibinfo{person}{Renjie Lv}, \bibinfo{person}{Kang Zhou}, {and} \bibinfo{person}{Yan Zhang}.} \bibinfo{year}{2023}\natexlab{}.
\newblock \showarticletitle{Is chatgpt a good recommender? a preliminary study}.
\newblock \bibinfo{journal}{\emph{arXiv preprint arXiv:2304.10149}} (\bibinfo{year}{2023}).
\newblock


\bibitem[Okura et~al\mbox{.}(2017)]%
        {okura2017embedding}
\bibfield{author}{\bibinfo{person}{Shumpei Okura}, \bibinfo{person}{Yukihiro Tagami}, \bibinfo{person}{Shingo Ono}, {and} \bibinfo{person}{Akira Tajima}.} \bibinfo{year}{2017}\natexlab{}.
\newblock \showarticletitle{Embedding-based news recommendation for millions of users}. In \bibinfo{booktitle}{\emph{Proceedings of the 23rd ACM SIGKDD international conference on knowledge discovery and data mining}}. \bibinfo{pages}{1933--1942}.
\newblock


\bibitem[Qi et~al\mbox{.}(2021)]%
        {qi2021pp}
\bibfield{author}{\bibinfo{person}{Tao Qi}, \bibinfo{person}{Fangzhao Wu}, \bibinfo{person}{Chuhan Wu}, {and} \bibinfo{person}{Yongfeng Huang}.} \bibinfo{year}{2021}\natexlab{}.
\newblock \showarticletitle{Pp-rec: News recommendation with personalized user interest and time-aware news popularity}.
\newblock \bibinfo{journal}{\emph{arXiv preprint arXiv:2106.01300}} (\bibinfo{year}{2021}).
\newblock


\bibitem[Qin et~al\mbox{.}(2023)]%
        {qin2023chatgpt}
\bibfield{author}{\bibinfo{person}{Chengwei Qin}, \bibinfo{person}{Aston Zhang}, \bibinfo{person}{Zhuosheng Zhang}, \bibinfo{person}{Jiaao Chen}, \bibinfo{person}{Michihiro Yasunaga}, {and} \bibinfo{person}{Diyi Yang}.} \bibinfo{year}{2023}\natexlab{}.
\newblock \showarticletitle{Is ChatGPT a general-purpose natural language processing task solver?}
\newblock \bibinfo{journal}{\emph{arXiv preprint arXiv:2302.06476}} (\bibinfo{year}{2023}).
\newblock


\bibitem[Radford et~al\mbox{.}(2018)]%
        {radford2018improving}
\bibfield{author}{\bibinfo{person}{Alec Radford}, \bibinfo{person}{Karthik Narasimhan}, \bibinfo{person}{Tim Salimans}, \bibinfo{person}{Ilya Sutskever}, {et~al\mbox{.}}} \bibinfo{year}{2018}\natexlab{}.
\newblock \showarticletitle{Improving language understanding by generative pre-training}.
\newblock  (\bibinfo{year}{2018}).
\newblock


\bibitem[Raffel et~al\mbox{.}(2020)]%
        {raffel2020exploring}
\bibfield{author}{\bibinfo{person}{Colin Raffel}, \bibinfo{person}{Noam Shazeer}, \bibinfo{person}{Adam Roberts}, \bibinfo{person}{Katherine Lee}, \bibinfo{person}{Sharan Narang}, \bibinfo{person}{Michael Matena}, \bibinfo{person}{Yanqi Zhou}, \bibinfo{person}{Wei Li}, {and} \bibinfo{person}{Peter~J Liu}.} \bibinfo{year}{2020}\natexlab{}.
\newblock \showarticletitle{Exploring the limits of transfer learning with a unified text-to-text transformer}.
\newblock \bibinfo{journal}{\emph{The Journal of Machine Learning Research}} \bibinfo{volume}{21}, \bibinfo{number}{1} (\bibinfo{year}{2020}), \bibinfo{pages}{5485--5551}.
\newblock


\bibitem[Staudemeyer and Morris(2019)]%
        {staudemeyer2019understanding}
\bibfield{author}{\bibinfo{person}{Ralf~C Staudemeyer} {and} \bibinfo{person}{Eric~Rothstein Morris}.} \bibinfo{year}{2019}\natexlab{}.
\newblock \showarticletitle{Understanding LSTM--a tutorial into long short-term memory recurrent neural networks}.
\newblock \bibinfo{journal}{\emph{arXiv preprint arXiv:1909.09586}} (\bibinfo{year}{2019}).
\newblock


\bibitem[Vaswani et~al\mbox{.}(2017)]%
        {vaswani2017attention}
\bibfield{author}{\bibinfo{person}{Ashish Vaswani}, \bibinfo{person}{Noam Shazeer}, \bibinfo{person}{Niki Parmar}, \bibinfo{person}{Jakob Uszkoreit}, \bibinfo{person}{Llion Jones}, \bibinfo{person}{Aidan~N Gomez}, \bibinfo{person}{{\L}ukasz Kaiser}, {and} \bibinfo{person}{Illia Polosukhin}.} \bibinfo{year}{2017}\natexlab{}.
\newblock \showarticletitle{Attention is all you need}.
\newblock \bibinfo{journal}{\emph{Advances in neural information processing systems}}  \bibinfo{volume}{30} (\bibinfo{year}{2017}).
\newblock


\bibitem[Wang et~al\mbox{.}(2023)]%
        {wang2023chatgpt}
\bibfield{author}{\bibinfo{person}{Zengzhi Wang}, \bibinfo{person}{Qiming Xie}, \bibinfo{person}{Zixiang Ding}, \bibinfo{person}{Yi Feng}, {and} \bibinfo{person}{Rui Xia}.} \bibinfo{year}{2023}\natexlab{}.
\newblock \showarticletitle{Is ChatGPT a good sentiment analyzer? A preliminary study}.
\newblock \bibinfo{journal}{\emph{arXiv preprint arXiv:2304.04339}} (\bibinfo{year}{2023}).
\newblock


\bibitem[Wu et~al\mbox{.}(2019a)]%
        {wu2019neural}
\bibfield{author}{\bibinfo{person}{Chuhan Wu}, \bibinfo{person}{Fangzhao Wu}, \bibinfo{person}{Mingxiao An}, \bibinfo{person}{Jianqiang Huang}, \bibinfo{person}{Yongfeng Huang}, {and} \bibinfo{person}{Xing Xie}.} \bibinfo{year}{2019}\natexlab{a}.
\newblock \showarticletitle{Neural news recommendation with attentive multi-view learning}.
\newblock \bibinfo{journal}{\emph{arXiv preprint arXiv:1907.05576}} (\bibinfo{year}{2019}).
\newblock


\bibitem[Wu et~al\mbox{.}(2019b)]%
        {wu2019neural2}
\bibfield{author}{\bibinfo{person}{Chuhan Wu}, \bibinfo{person}{Fangzhao Wu}, \bibinfo{person}{Suyu Ge}, \bibinfo{person}{Tao Qi}, \bibinfo{person}{Yongfeng Huang}, {and} \bibinfo{person}{Xing Xie}.} \bibinfo{year}{2019}\natexlab{b}.
\newblock \showarticletitle{Neural news recommendation with multi-head self-attention}. In \bibinfo{booktitle}{\emph{Proceedings of the 2019 conference on empirical methods in natural language processing and the 9th international joint conference on natural language processing (EMNLP-IJCNLP)}}. \bibinfo{pages}{6389--6394}.
\newblock


\bibitem[Wu et~al\mbox{.}(2022)]%
        {wu2022news}
\bibfield{author}{\bibinfo{person}{Chuhan Wu}, \bibinfo{person}{Fangzhao Wu}, \bibinfo{person}{Tao Qi}, \bibinfo{person}{Chenliang Li}, {and} \bibinfo{person}{Yongfeng Huang}.} \bibinfo{year}{2022}\natexlab{}.
\newblock \showarticletitle{Is News Recommendation a Sequential Recommendation Task?}. In \bibinfo{booktitle}{\emph{Proceedings of the 45th International ACM SIGIR Conference on Research and Development in Information Retrieval}}. \bibinfo{pages}{2382--2386}.
\newblock


\bibitem[Wu et~al\mbox{.}(2020)]%
        {wu2020mind}
\bibfield{author}{\bibinfo{person}{Fangzhao Wu}, \bibinfo{person}{Ying Qiao}, \bibinfo{person}{Jiun-Hung Chen}, \bibinfo{person}{Chuhan Wu}, \bibinfo{person}{Tao Qi}, \bibinfo{person}{Jianxun Lian}, \bibinfo{person}{Danyang Liu}, \bibinfo{person}{Xing Xie}, \bibinfo{person}{Jianfeng Gao}, \bibinfo{person}{Winnie Wu}, {et~al\mbox{.}}} \bibinfo{year}{2020}\natexlab{}.
\newblock \showarticletitle{Mind: A large-scale dataset for news recommendation}. In \bibinfo{booktitle}{\emph{Proceedings of the 58th Annual Meeting of the Association for Computational Linguistics}}. \bibinfo{pages}{3597--3606}.
\newblock


\bibitem[Xu et~al\mbox{.}(2023)]%
        {xu2023openp5}
\bibfield{author}{\bibinfo{person}{Shuyuan Xu}, \bibinfo{person}{Wenyue Hua}, {and} \bibinfo{person}{Yongfeng Zhang}.} \bibinfo{year}{2023}\natexlab{}.
\newblock \showarticletitle{OpenP5: Benchmarking Foundation Models for Recommendation}.
\newblock \bibinfo{journal}{\emph{arXiv:2306.11134}} (\bibinfo{year}{2023}).
\newblock


\bibitem[Yang(2016)]%
        {yang2016effects}
\bibfield{author}{\bibinfo{person}{JungAe Yang}.} \bibinfo{year}{2016}\natexlab{}.
\newblock \showarticletitle{Effects of popularity-based news recommendations (“most-viewed”) on users' exposure to online news}.
\newblock \bibinfo{journal}{\emph{Media Psychology}} \bibinfo{volume}{19}, \bibinfo{number}{2} (\bibinfo{year}{2016}), \bibinfo{pages}{243--271}.
\newblock


\bibitem[Zhang et~al\mbox{.}(2023)]%
        {zhang2023chatgpt}
\bibfield{author}{\bibinfo{person}{Jizhi Zhang}, \bibinfo{person}{Keqin Bao}, \bibinfo{person}{Yang Zhang}, \bibinfo{person}{Wenjie Wang}, \bibinfo{person}{Fuli Feng}, {and} \bibinfo{person}{Xiangnan He}.} \bibinfo{year}{2023}\natexlab{}.
\newblock \showarticletitle{Is chatgpt fair for recommendation? evaluating fairness in large language model recommendation}.
\newblock \bibinfo{journal}{\emph{arXiv preprint arXiv:2305.07609}} (\bibinfo{year}{2023}).
\newblock


\bibitem[Zhang and Wang(2023)]%
        {zhang2023prompt}
\bibfield{author}{\bibinfo{person}{Zizhuo Zhang} {and} \bibinfo{person}{Bang Wang}.} \bibinfo{year}{2023}\natexlab{}.
\newblock \showarticletitle{Prompt learning for news recommendation}.
\newblock \bibinfo{journal}{\emph{arXiv preprint arXiv:2304.05263}} (\bibinfo{year}{2023}).
\newblock


\end{thebibliography}
\end{document}